\documentclass[prl,twocolumn,nofootinbib]{revtex4}
\usepackage{graphicx}
\usepackage{latexsym}
\def\be{\begin{equation}}
\def\ee{\end{equation}}
\def\bea{\begin{eqnarray}}
\def\eea{\end{eqnarray}}

\begin{document}
\title{Reconstructing large running-index inflaton potentials}
\author{Chiyi Chen${}^{a,b}$}
\author{Bo Feng${}^c$}
\author{Xiulian Wang${}^{d}$}
\author{Zhaoyu Yang${}^{c}$}
\affiliation{${}^a$Shanghai Astronomical Observatory, Chinese
Academy of Sciences, Shanghai 200030, PR China}
\affiliation{${}^b$National Astronomical Observatories, Chinese
Academy of Sciences, Beijing 100012, PR China}
\affiliation{${}^c$Institute of High Energy Physics, Chinese
Academy of Science, P.O. Box 918-4, Beijing 100039, P. R. China}
\affiliation{${}^d$Institute of Theoretical Physics, Chinese
Academy of Sciences, Beijing 100080, P. R. China.}

\begin{abstract}
Recent fits of cosmological parameters by the first year Wilkinson
Microwave Anisotropy Probe (WMAP) measurement seem to favor a
primordial scalar spectrum with a large varying index from blue to
red. We use the inflationary flow equations to reconstruct large
running-index inflaton potentials and comment on current status on
the inflationary flow. We find previous negligence of higher order
slow rolling contributions when using the flow equations
would lead to  unprecise results.  \\
PACS number(s): 98.80.Cq, 11.10.Kk
\end{abstract}

\maketitle In the past decade, inflation theory has successfully
passed several nontrivial tests. In particular, the recently
released Wilkinson Microwave Anisotropy Probe (WMAP) data
\cite{kogut} have detected a large-angle anti-correlation in the
temperature--polarization cross-power spectrum, which is the
signature of adiabatic superhorizon fluctuations at the time of
decoupling\cite{WMAP2}. The high-precision WMAP data have been
used to fit the cosmological parameters and confront the
predictions of inflationary scenarios respectively in
Refs.~\onlinecite{WMAP} and \onlinecite{WMAP2}. It is noted that
there might be possible discrepancies between predictions and
observations on the largest and smallest scales. Basing on this
fact Spergel \cite{WMAP} et al introduced a new parameter:
$dn_s/d\ln k$ in the fit. With the global fit to WMAPext, 2df
\cite{2df} and Lyman-$\alpha$ forests \cite{forest}the authors
found a preference of nonzero running at around $2\sigma$.
Although the use of Lyman-$\alpha$ data was questioned by
\cite{Seljak}, the fit to WMAP alone favored the running to about
$1.5\sigma$: at the pivot scale $k_0=0.002~{\rm Mpc}^{-1}$, the
best-fit values of the scalar power spectrum for WMAP alone being
$n_s=1.20^{+0.12}_{-0.11}$ and $dn_s/d\ln
k=-0.077^{+0.050}_{-0.052}$ \cite{WMAP2}. It is recently found
that the inclusion of VSA data can lead to a negative running at a
level of more than $95\%$ confidence when fitting to CMB
only\cite{VSA}. Theoretically there have been studies in the
literature since the release of the WMAP data on models of
inflation which provide a running index required by the WMAP
\cite{feng,running}. On the other hand, Bridle et al claimed that
the need of running should come from the low CMB quadrupole
alone\cite{Lewis} and thereafter many theoretical models were put
forward\cite{smalll}.  Meanwhile Mukherjee and Wang \cite{Wang}
used the model independent reconstruction of primordial spectrum
and found that a running index from blue to red was indeed
favored.

The need of large running has been studied
widely\cite{Wang,Lewis,Seljak,fitWMAP,Kinney03} after the release
of first year WMAP data. At least WMAP could not rule out
significant running and if further stands, it could severely
constrain inflation model buildings. In the present paper we will
use the inflationary flow equations to generate potentials which
can lead to large variations in the spectral index.

The inflationary flow equations were firstly introduced by Hoffman
and Turner\cite{HT01} to study generic predictions of slow-rolling
inflation and for the first time compared with observations in
Ref.\onlinecite{kunz}. Latter Kinney\cite{Kinney02}developed the
equations as the inflaton potential ``generator", which has been
used widely in the literature\cite{WMAP2,Dodelson,Kinney03}. It
was later shown by Ref.\cite{Ke03} that the flow equations could
be used to reconstruct inflaton potentials.  By randomly
generating a large number of initial flow parameters, evolving
towards the observed scale and comparing with the observations,
inflaton potentials that satisfy the observations are then
reconstructed\cite{Ke03}. Thereafter Liddle\cite{liddle03flow}
pointed out subtly some disadvantages with current version of
inflationary flow. We will reconstruct inflaton potentials with
large running of spectral index, comment on Liddle's remarks and
put forward our understandings on inflationary flow.

We first follow closely the notations by Kinney\cite{Kinney02}.
For a single scalar field  inflaton  $\phi$ it obeys the equation
of motion
\begin{equation}
\ddot \phi + 3 H \dot \phi + V'(\phi) =
0,\label{eqequationofmotion}
\end{equation}
where $H \equiv (\dot a / a)$ is the Hubble parameter and
$V(\phi)$ is its potential. Another equation is the   {\it
Hamilton-Jacobi} equation
\begin{equation}
H^2(\phi) \left[1 - {1\over 3} \epsilon(\phi)\right] =  \left({8
\pi \over 3 m_{\rm Pl}^2}\right)
 V(\phi),\label{eqHJ}
\end{equation}
where
\begin{equation}
\epsilon \equiv {m_{\rm Pl}^2 \over 4 \pi} \left({H'(\phi) \over
 H(\phi)}\right)^2.\label{eqepsilon}
\end{equation}

The scale factor during inflation is given by
\begin{equation}
a \propto e^{N} = \exp\left[\int_{t_0}^{t}{H\,dt}\right],
\end{equation}
where the number of e-folds $N$ is
\begin{equation}
N \equiv \int_{t}^{t_e}{H\,dt} = \int_{\phi}^{\phi_e}{{H \over
\dot\phi}\,d\phi} = {2 \sqrt{\pi} \over m_{\rm Pl}}
\int_{\phi_e}^{\phi}{d\phi \over \sqrt{\epsilon(\phi)}}.
\end{equation}

The slow-roll parameter $\epsilon$ can be expressed by
\begin{equation}
\sqrt{\epsilon} = {m_{\rm Pl} \over 2 \sqrt{\pi}} {H' \over H}
\end{equation}
in the convention of Hubble expansion. Accordingly, high order
slow-roll  parameters in Hubble expansion can be given
\cite{liddle94}:
\begin{eqnarray}
\sigma &\equiv& {m_{\rm Pl} \over \pi} \left[{1 \over 2}
\left({H'' \over
 H}\right) -
\left({H' \over H}\right)^2\right],\cr {}^\ell\lambda_{\rm H}
&\equiv& \left({m_{\rm Pl}^2 \over 4 \pi}\right)^\ell
{\left(H'\right)^{\ell-1} \over H^\ell} {d^{(\ell+1)} H \over
d\phi^{(\ell + 1)}}.
\end{eqnarray}
Using the equation
\begin{equation}
{d \over d N} = {d \over d\ln a} = { m_{\rm Pl} \over 2
\sqrt{\pi}} \sqrt{\epsilon} {d \over d\phi},
\end{equation}
it is convenient to take the derivative with respect to number of
e-folds instead of $\phi$. The evolution of above slow-roll
parameters can be described by the ``flow'' equations
\cite{HT01,Kinney02}
\begin{eqnarray}
{d \epsilon \over d N} &=& \epsilon \left(\sigma + 2
\epsilon\right),\cr {d \sigma \over d N} &=& - 5 \epsilon \sigma -
12 \epsilon^2 + 2 \left({}^2\lambda_{\rm H}\right),\cr {d
\left({}^\ell\lambda_{\rm H}\right) \over d N} &=& \left[
\frac{\ell - 1}{2} \sigma + \left(\ell - 2\right) \epsilon\right]
\left({}^\ell\lambda_{\rm H}\right) + {}^{\ell+1}\lambda_{\rm
H}.\label{eqfullflowequations}
\end{eqnarray}
As shown by Ref.\onlinecite{Kinney02} the flow equations have a
stable late time attractor with
\begin{eqnarray}
\epsilon &=& {}^\ell\lambda_{\rm H} = 0,\cr \sigma &=& {\rm
const.}\label{eqfixedpoint1}
\end{eqnarray}
  For any single field inflation model
with common dynamics (i.e., satisfying above equation of motion
and Hamilton-Jacobi equation ), the background dynamics of the
Hubble parameter (up to a normalization) can be equivalently
described by flow equations as long as $l$ can be extended to
infinity. However, numerical calculation cannot accommodate
infinite set of equations and one has to truncate above flow
equations. A common truncation to $M-th$ order assumes the
${}^\ell\lambda_{\rm H}$ are all zero for $l>M$. After truncation
the flow equations cannot accommodate $all$ the single field
inflation models, as shown by Liddle\cite{liddle03flow}, but the
flow equation itself is still $exact$. We have
\begin{equation}
{1 \over H} {d H \over d N} = \epsilon,  ~~~~~  {d \phi \over d N}
= {m_{\rm Pl} \over 2 \sqrt{\pi}} \sqrt{\epsilon}.
\end{equation}
For given initial values of $\phi$ and $H$, their trajectories can
be correspondingly given via the flow equations, $V(\phi)$ can
then be worked out with the Hamilton-Jacobi equation once $H$ is
known. The initial value of $\phi$ is arbitrary  and one can set
it to be zero, but $H$ has to be determined on the observable
scales from the primordial scalar spectrum (e.g., from WMAP
\cite{WMAP2} ) :
\begin{equation}
\mathcal{P}_R \approx {H^2 \over \pi \epsilon m^2_{\rm Pl}}
\approx 2.2 \times 10^{-9}.
\end{equation}

 The inflationary ``observables'' include
tensor/scalar ratio $r$, the spectral index $n_S$, and the
``running'' of the spectral index $d n_S / d\ln k$, etc. To the
second order in slow roll(SR) one gets \cite{liddle94,stewart93},
\begin{equation}
r = 16 \epsilon \left[1 - C \left(\sigma + 2
 \epsilon\right)\right],\label{eqrsecondorder}
\end{equation}
for the tensor/scalar ratio, and
\begin{equation}
n_S - 1 = \sigma - \left(5 - 3 C\right) \epsilon^2 - {1 \over 4}
\left(3 - 5 C\right) \sigma \epsilon + {1 \over 2}\left(3 -
C\right) \left({}^2\lambda_{\rm H}\right)\label{ns2}
\end{equation}
for the spectral index. $C \equiv 4 (\ln{2} + \gamma)- 5 $, where
$\gamma \simeq 0.577$ is Euler's constant. $dn_S/d\ln k$ can be
 given via the relation
\begin{equation}
{d n_S \over d \ln k} = - \left({1 \over 1 - \epsilon}\right) {d n
\over d N},
\end{equation}
which is tedious but can be directly worked out from
Eq.~(\ref{ns2}). The number of e-folds $N$  which corresponds to
the observable scale is largely uncertain due to the uncertainty
in the energy density during inflation and the reheating
temperature \cite{lidsey95,fgw03}, typically in the range [40,60].

Despite the doubt on whether flow equations correspond directly to
inflationary dynamics, we can reconstruct inflaton potentials
which lead to large variations in the spectral index using the
Monte Carlo method, which may be still exact after a truncation at
higher order in the flow equations. Similar to
Ref.\onlinecite{Ke03}, the algorithm for our Monte Carlo
reconstruction is as follows:
\begin{enumerate}
\item Specify a ``window'' of parameter space of the observables:
e.g. 1$\sigma$ WMAP constraints on $n_S-1$, $r$ and $d n_S /d
\ln{k}$ and their associated error bars.
\item Select a random point in slow roll space,
$[\epsilon,\delta,{}^\ell\lambda_{\rm H}]$, truncated at order $M$
in the slow roll expansion.

\item Evolve forward in time ($d N < 0$) until either (a) inflation ends
 ($\epsilon > 1$), or (b) the evolution reaches a late-time fixed
 point ($\epsilon = {}^\ell\lambda_{\rm H} = 0,\ \sigma = {\rm
 const}$), or (c) $|{}^\ell\lambda_{\rm H}|>100$, or (d) inflation does
 not end after evolving $\Delta N >1000$ .
\item If the evolution reaches a late-time fixed point, goto 6
(Our main intention is to search large $d n_S /d \ln{k}$, which is
zero is this case ). If the evolution reaches
$|{}^\ell\lambda_{\rm H}|>100$, for the sake of saving computing
time and to ensure the validity of expanding the observables to
second order in slow roll, goto 6. If inflation does
 not end after evolving $\Delta N >1000$, record it as
 insignificant point, goto 6.
\item If inflation ends, evolve the flow equations backward 40
until 55 e-folds from the end of inflation. Calculate the
observable parameters at  each point. Once the observable window
is satisfied, record it as nontrivial point, compute the potential
and add this model to the ensemble of ``reconstructed'' potentials
and go to 6 . Else evolve until 55, calculate the observables and
record as a nontrivial point.

\item Repeat steps 2 to 5 until the desired number of models
have been studied.
\end{enumerate}

We truncate the slow roll hierarchy to 5th order, the model
parameters are randomly from the following uniform distributions:
\begin{eqnarray}
\epsilon &=& \left[0,0.8\right]\cr \sigma &=&
\left[-0.5,0.5\right]\cr {}^2\lambda_{\rm H} &=&
\left[-0.05,0.05\right]\cr {}^3\lambda_{\rm H} &=&
\left[-0.025,0.025\right],\cr &\cdots&\cr {}^{6}\lambda_{\rm H}
&=& 0.\label{eqinitialconditions}
\end{eqnarray}
The thinning on the initial slow roll hierarchy is set to ensure a
reasonable convergence. However it is still possible that one gets
larger values for higher order parameters, so we set a ``lock''
that $|{}^\ell\lambda_{\rm H}|\leq 100$ in our simulation. As we
are merely intending to find several potentials which can give
large $d n_S /d \ln{k}$, our method is in this sense applicable.
For a 1000,000 simulation, we get
\begin{itemize}
\item{Late-time attractor: 921,793.}
\item{Nontrivial points: 75,300.}
\item{$|{}^\ell\lambda_{\rm H}|>100$: 2,906.}
\item{Insignificant: 1.}
\end{itemize}

A companion running of truncation to 8th order is also tried for
crosscheck. In this case we use almost the same method as the WMAP
team \cite{WMAP2} except the negligence of $|{}^\ell\lambda_{\rm
H}|>10^8$(this is taken as Method Crosscheck below). This time we
set the range $N= \left[40,70\right]$, we get (for a 1000,000
simulation)
\begin{itemize}
\item{Late-time attractor: 924,164.}
\item{Nontrivial points: 75,646.}
\item{$|{}^\ell\lambda_{\rm H}|>10^8$: 187.}
\item{Insignificant: 3.}
\end{itemize}

 We plot our zoo of the nontrivial points in
Fig.1. Interestingly there are some straight lines for $n \sim
1.01$ and $d n_S /d \ln{k}\sim -0.02$, we shall make further
comments on this below. Meanwhile the strait lines are unavailable
for Method Crosscheck, as shown in Fig.2. The ``window'' we open
here is $around$ 1$\sigma$ WMAP constraints at $k=0.002$
$Mpc^{-1}$: $1.01<n_S<1.4$, $r<1.14$, $-0.021<d n_S /d
\ln{k}<-0.13$. For the primordial spectrum with large running in
the spectral index, it is extremely difficult to make an exact
description on the resulted spectrum: WMAP team assumed a constant
running on spectral index in their fit and gave the above
1$\sigma$ region. However from the evolution of flow equations $d
n_S /d \ln{k}$ is generically not a constant. Under such
circumstances many wrong conclusions would be reached once only
above window is used. Firstly small nonzero $d(d n_S /d \ln{k})/d
\ln{k}$ other than the assumption taken by WMAP team is no doubt a
possibly good fit to WMAP, the observational data can never
restrict higher order terms to be exactly zero. Secondly what WMAP
constraints on is a continuous region from $10^{-4} \sim 6 \times
10^{-2}$ h$Mpc^{-1}$, the parameters from flow may change
significantly within few number of e-folds, this may lead to some
obvious mismatching. Theoretically flow equations can even be used
to rebuild potentials which lead to suppressed large scale
primordial spectrum or those which get a kick on smaller scales to
achieve larger CMB TE multipoles on the largest scales. But the
window is extremely difficult to set. In the case of a large $d
n_S /d \ln{k}$ we have to make some evaluations by hand. The first
step is that we open the ``window'' as stated above, we get 858
points which appear in the ``window'' for the 1000,000 iterations
and 8,788 points for a 10,000,000 simulation. We plot the 8,788
``raw'' data in Fig.3. We make a hand selection around the first
100 ``raw'' points, the data with global blue $n$ and smaller
running have been extracted.

After careful extracting we find 81 of the data lead to global
blue index for $N>30$ (when reanalyzing $N$ is loosened to 30,
which is still acceptable respecting the reheating temperature
limit ), 5 where  the constraint on $n_S$ and $d n_S /d \ln{k}$
cannot be simultaneously satisfied, 2 where second order SR does
not work well (for detailed discussions see below), only 12 points
are left, leading to running enough from blue to red (It is
theoretically plausible to exclude those which lead to global blue
index, which could be added to the window). We show the 12
resulting $n_S$ in Fig.4, their initial flow parameters are shown
in Table 1. The corresponding potentials are shown in Fig.5 and
their trajectories in Fig.6. The trajectories are rather
complicated. Very interestingly some trajectories overlap during
some period in Fig.6, as can also be seen from Fig.5, around the
CMB and Large Scale Structure interest scale(black/dark lines)
some potentials do overlap, but are then separated on other
scales. We find that $r$ is no more than $0.5$ for the twelve
points. Similarly, if necessary, one can work on all the 8,788
``raw'' data and filtering out hundreds of potentials that lead to
large variations in the index.
\begin{figure}[htbp]
\begin{center}
\includegraphics[scale=0.70]{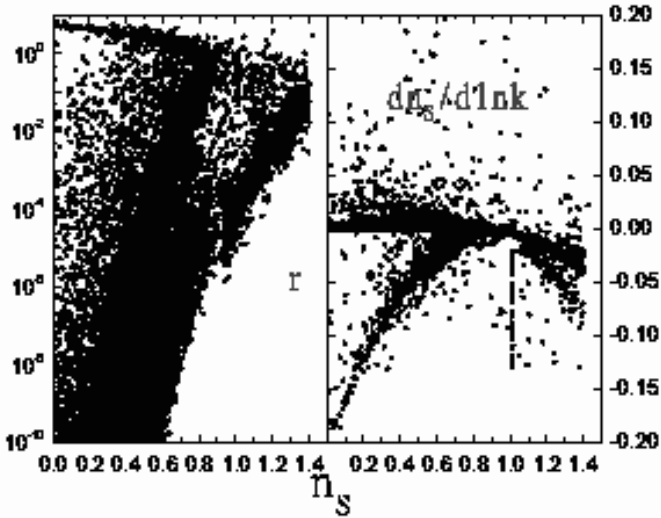}
\caption{ Models plotted in the $(n_S,r)$ and $(n_S,d n_S /d
\ln{k})$ plane for an $M=5$ Monte Carlo. The cause of short
straight lines around $n_S=1$ and $d n_S /d \ln{k}= -0.02$ can be
found in Fig.7. \label{fig:fig1}}
\end{center}
\end{figure}

\begin{figure}[htbp]
\begin{center}
\includegraphics[scale=0.70]{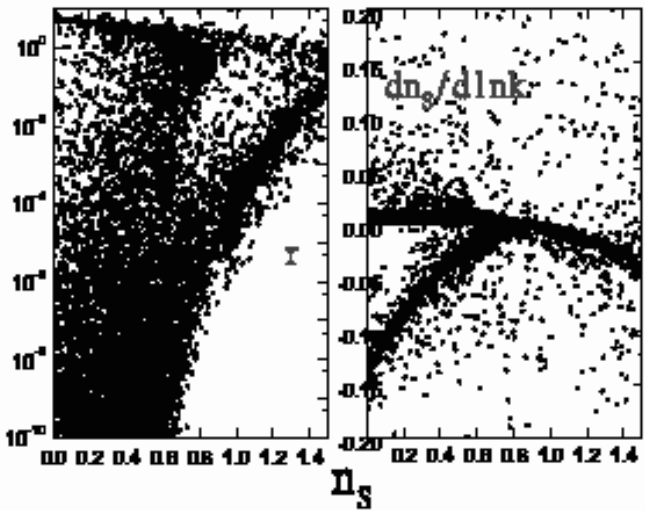}
\caption{ Models plotted in the $(n_S,r)$ and $(n_S,d n_S /d
\ln{k})$ plane for an $M=8$ Monte Carlo. The number of e-folds is
now randomly generated in range $\left[40,70\right]$.}
\end{center}
\end{figure}


\begin{figure}[htbp]
\begin{center}
\includegraphics[scale=0.35]{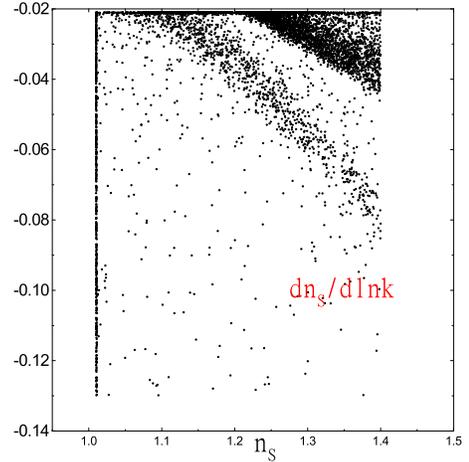}
\caption{ Models plotted in the $(n_S,d n_S /d \ln{k})$ plane for
a 10,000,000 simulation fitting into the window around 1$\sigma$
WMAP constraints at $k=0.002$ $Mpc^{-1}$. The cause of the
straight line around $n_S=1$ can be found in Fig.7.
\label{fig:fig3}}
\end{center}
\end{figure}

\begin{figure}[htbp]
\begin{center}
\includegraphics[scale=0.35]{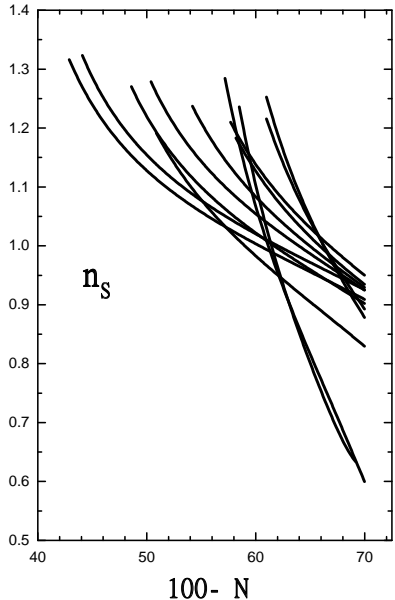}
\caption{ Twelve resulting spectral indices that runs
significantly from blue to red, satisfying 1$\sigma$ WMAP
constraints\cite{WMAP2}. \label{fig:fig4}}
\end{center}
\end{figure}

\begin{figure}[htbp]
\begin{center}
\includegraphics[scale=0.35]{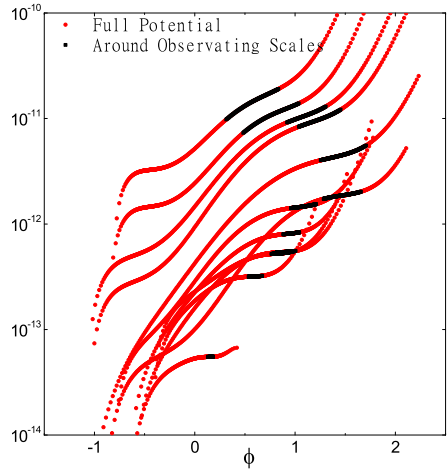}
\caption{ Twelve reconstructed inflaton potentials from Fig.4.
\label{fig:fig5}}
\end{center}
\end{figure}

\begin{figure}[htbp]
\begin{center}
\includegraphics[scale=0.35]{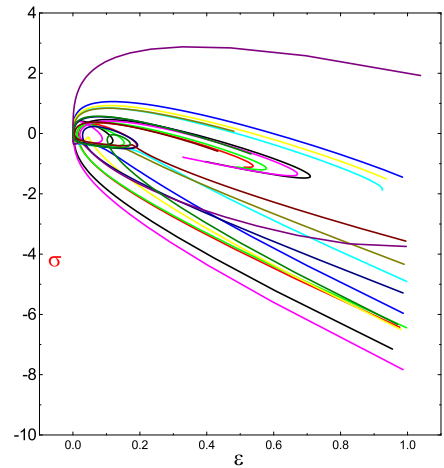}
\caption{ Twelve trajectories of $\epsilon$-$\sigma$ from Fig.4
during inflation.}
\end{center}
\end{figure}

\begin{figure}[htbp]
\begin{center}
\includegraphics[scale=0.6]{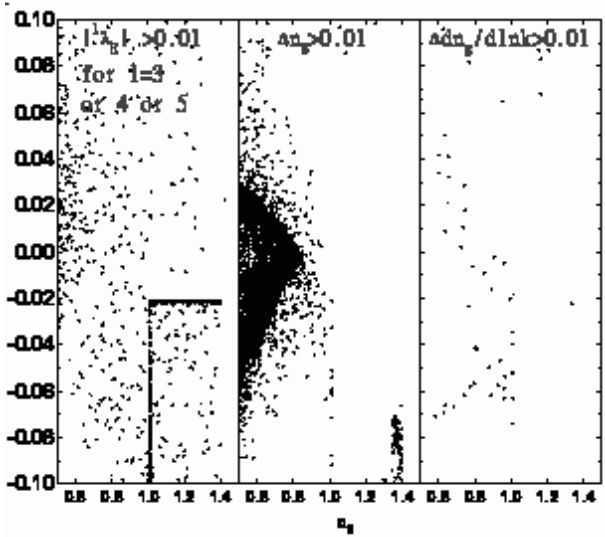}
\caption{Crosscheck of SR approximations for a 10,000,000
simulation on the observable values of $n_S$ and $dn_S/d\ln k$,
with $dn_S/d\ln k$ being the vertical axis. The left panel shows
those which lead to wrong observable values due to nonegligible
higher order effects of ${}^l\lambda_{\rm H}$ ($l\geq 3$); The
middle and right panel show those which give wrong values of $n_S$
and $dn_S/d\ln k$ due to insufficient SR approximation to second
order.}
\end{center}
\end{figure}

\begin{figure}[htbp]
\begin{center}
\includegraphics[scale=0.65]{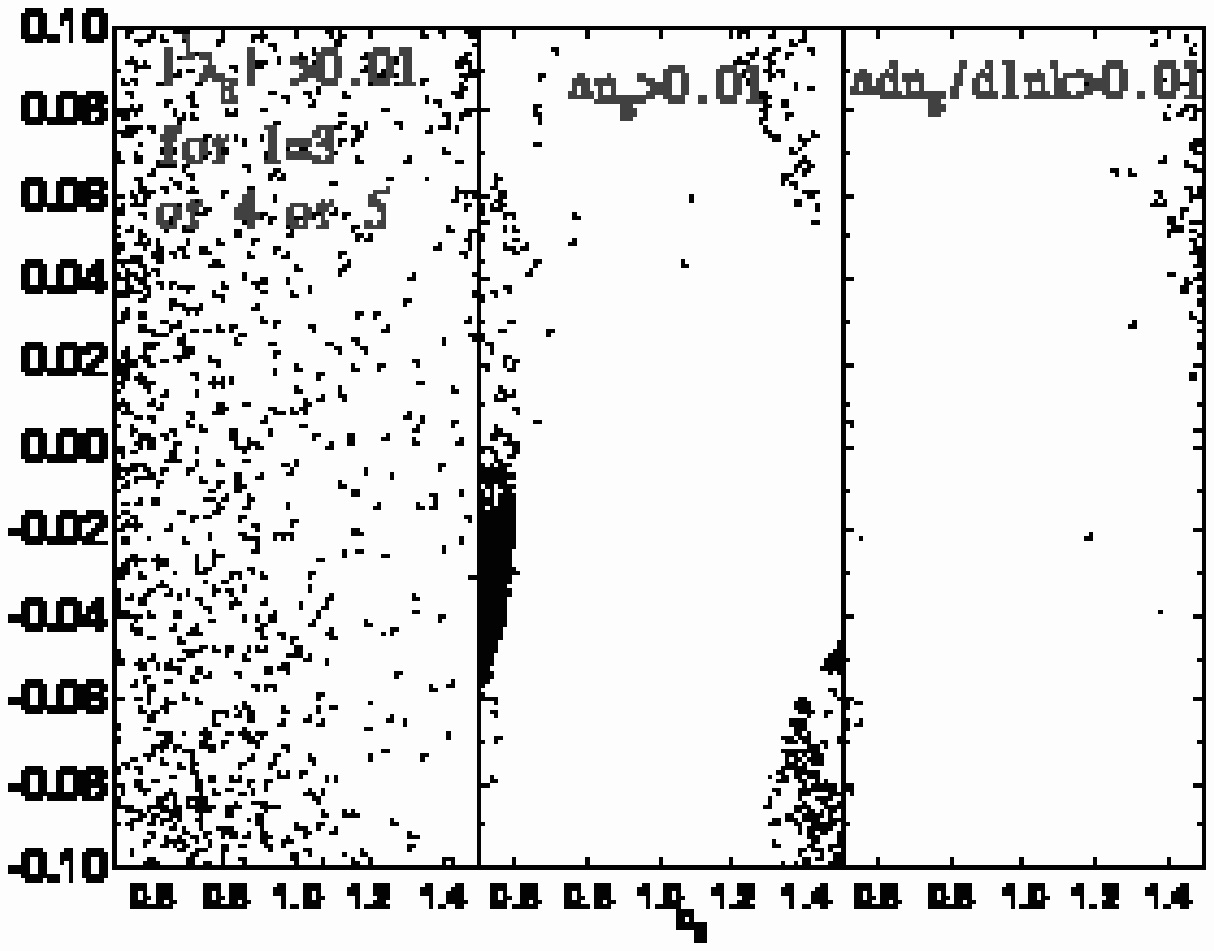}
\caption{The same as Fig.7 for a 10,000,000 iteration on Method
Crosscheck described in the text.}
\end{center}
\end{figure}

 There have been in the literature several works
which try to reconstruct inflaton potential using ideal
observational data\cite{ckll03,lidsey95,Ke03}. Generally speaking,
first year WMAP data have not provided stringent window for one to
constrain the inflaton potential. WMAP has not detected tensor
contributions in the primordial spectrum: $r\leq 1$\cite{WMAP2}.
This leads to $\epsilon \leq 0.07$ and hence $H^2\leq 4.8 \times
10^{-10} m^2_{Pl}$, this gives $0<V \leq 5.8 \times 10^{-11}
m^4_{Pl} $ via the Hamilton-Jacobi equation. As also shown
although a large running is favored, WMAP is consistent with scale
invariant spectrum. Peiris et.al.\cite{WMAP2} categorized the slow
roll parameters into four classes for generic single field
inflaton potentials and found for either class there is broad
region consistent with current observations. In this sense first
year WMAP data (together with other observational data) have not
been able to work as a stringent enough discriminator. As for the
reconstructing method, there would be broad parametric space when
the observational window is less stringent. However as the
observational data accumulate and the error bar shrinks, e.g. when
tensor contribution is exactly measured and running of the
spectral index is strongly confirmed, reconstructing the favored
inflaton potential would make its sense. On the inflationary flow
itself, it would be inefficient since very small fraction fall
into our window above, meanwhile most of the iterations fall into
lat-time attractors and little part satisfies the window. In any
case it provides a testable inflaton potential generator which may
be exactly solvable. Liddle\cite{liddle03flow} has made some
detailed descriptions on its shortcomings: Firstly the flow
equations (Eq.9 in this paper) do not relate directly to the
inflationary dynamics. This should not be a severe problem since
the reconstruction of potential needs using the Hamilton-Jacobi
equation, where the dynamics is included. Secondly due to the
truncation on $M-th$ order, the effective potential is only in
such form
\begin{eqnarray}
\label{e:Vsol} V(\phi) &=& \frac{3 m_{{\rm Pl}}^2}{8\pi} \, H_0^2
\left(1+A_1 \phi + \cdots +
A_{M+1} \phi^{M+1} \right)^2   \\
 && \hspace*{-26pt} \times\left[ 1 - \frac{1}{3} \frac{m_{{\rm Pl}}^2}{4\pi}
\left( \frac{A_1 + \cdots + (M+1)A_{M+1} \phi^M}{1+A_1\phi +
\cdots + A_{M+1}\phi^{M+1}} \right)^2 \right]
\,.\nonumber\label{liddlev}
\end{eqnarray}
We made a detailed check and found the form is satisfied exactly.
This in fact does accomodate many inflaton potentials as there are
so many undecided parameters to choose and  an ideally no
truncation would accomodate all the single field inflaton
potentials which satisfy above equation of motion and the
Hamilton-Jacobi equation. A main loophole is pointed by Liddle
\cite{liddle03flow} that for the nontrivial points where inflation
ends at $\epsilon =1$ the full potential is negative at its
minimum. For example Eq.17 reduces to $V(\phi)=V_0 ((1+A_1\phi)^2-
\frac{A_1}{12 \pi})$ when only $\epsilon$ is nonzero where the
flow equation starts to evolve (we set $\phi$ to be zero here).
However as the flow equation only describes the dynamics of
inflation at $\epsilon \leq 1$ where $V$ is always positive (as
can be clearly seen from Eq.2 ), we can assume that the exact
potential after inflation can be matched by other potentials which
reach the minimum at $V=0$ (or arguably matched at $\epsilon=1$),
the perturbations would certainly be the same as in Eq.17. In this
sense flow equations can reconstruct potentials $during$
inflation. Similar arguments hold for the fixed points where
inflaton reaches a local positive minimum and drives eternal
inflation, this situation can be related to the standard hybrid
inflation mechanism\cite{linde91,linde94}, as shown by Easther and
Kinney\cite{Ke03}. In any case the flow equations seem to be
inefficient as the inflaton potential generator when confronted
with a large negative running from blue to red, which is around
the 1$\sigma$ region of first year data of WMAP. Further
investigation to find better infaton generator is necessary
confronted with future precise observations.

So far we have left the second order SR approximation(SRA)
undiscussed (We thank the anonymous referees for inspirations on
this issue). It has been noted in the literature\cite{horder} that
higher order SR contributions may not be negligible especially
when confronted with high precision observational data. This being
the first time to check the precision of SR approximation in the
framework of inflationary flow, we first rewrite the spectral
index to the third order in the form of flow parameters(Basing on
Refs.\onlinecite{GStewart,STG01}):
\begin{eqnarray}
n_S - 1 & = & \sigma +2 \alpha {}^2\lambda_{\rm H} + \left(
\alpha^2 - \frac{\pi^2}{12} \right) {}^3\lambda_{\rm H} + \left(
4-12\alpha
\right) \epsilon^2  \nonumber \\
&  & + \left( 3-5\alpha\right) \epsilon \sigma + \left(-5 \alpha^2
+8 \alpha -6 + \frac{5 \pi^2}{ 12} \right) \epsilon ~
{}^2\lambda_{\rm H} \nonumber\\
& & + \left( \frac{\alpha^2}{2} +4 - \frac{13 \pi^2}{24} \right)
\sigma ~{}^2\lambda_{\rm H} \nonumber\\
& & + \left( 6 \alpha^2 -32 \alpha +28 -\frac{\pi^2}{2} \right)
\epsilon^3 \nonumber\\
& & + \left( -\frac{9}{2} \alpha^2 -6 \alpha -18 + \frac{27 }{8}
\pi^2 \right) \epsilon^2 \sigma \nonumber\\
& & + \left( -\frac{5}{2} \alpha^2 +3 \alpha -13 +\frac{35}{24}
\pi^2 \right) \epsilon \sigma^2 \label{ns3}
\end{eqnarray}
where $\alpha =\frac{3-C}{4} =0.7296$. We made three sets of
crosscheck in our 1000,000 and 10,000,000 iterations. Firstly one
can see from Eq.18 the factors on ${}^2\lambda_{\rm H}$ and
${}^3\lambda_{\rm H}$ differ no more than five times. There being
no efforts available in higher order SRA as
Ref.\onlinecite{GStewart} has explicitly did, we first make a
strong limit that $max($ $\mid{}^3\lambda_{\rm H}\mid$,
$\mid{}^4\lambda_{\rm H}\mid$, $\mid{}^5\lambda_{\rm H} \mid) <
0.01$ (taken as Condition A)is satisfied. Those which violate this
condition are plotted in the left panel of Fig.7. Other limits are
taken to ensure that the values of $d n_S/d \ln k$ (taken as
Condition B)and $n_S$ (taken as Condition C)differ no more than
0.01 when a third order SRA is assumed instead of second order
SRA. They are delineated in the right and middle panels of Fig.7.
To our great surprise 319, 134 and 6376 points violate Conditions
A,B and C respectively in the 1000,000 iteration and for the
10,000,000 iteration 3089, 1404 and 64589 points violate the three
conditions ( 751,215 total nontrivial points for the 10,000,000
iteration). That is to say, near ten percents of nontrivial points
need to be reconsidered when a third order SRA is taken instead of
second order SRA. A natural conclusion is that one has to take the
order of SRA as high as possible to ensure the validity of SRA. It
seems that we are lucky enough today- as can be seen from Fig.7-
the regions where Conditions B and C are violated seems to be
disfavored by current observations. For the stringent condition A
we can find the straight lines in Figs.1,3 are due to the
violation of this condition. It is however noted that Conditions B
and C are the weak conditions which work only to third order and
nothing could be ensured under fourth or fifth order SRA. Would
this be the real case, one has to solve Eq.17 mode by mode
instead. (A naive way out $might$ be  to ensure the validity of
decreasing sufficiently the higher order flow parameters.)  It is
worth mentioning again that the flow equations of Eq.9 is always
exact under any order of truncation. In this sense the way of
using flow equations like Ref.\onlinecite{Dodelson} is
appropriate, while many other papers in the literature suffer from
second order SRA\cite{Kinney02,Ke03,WMAP2,Kinney03,probflow}. When
a truncation to eighth order is considered instead we find similar
results. For an iteration of 1000,000 points with truncation to
the eighth order(Method Crosscheck) shown in Fig.2 we get 1,972,
289 and 5,216 points for models that violate Conditions A,B and C.
A 10,000,000 iteration of Method Crosscheck is also tried and we
get 17 insignificant points, 1,841 with $|{}^\ell\lambda_{\rm
H}|>10^8$ and 757,521 nontrivial points. The 19,719, 2,804 and
51,626 points which violate Conditions A, B and C are shown in
corresponding window in Fig.8. \footnote{We've made some changes
on the original Figs.1,2,7,8 to take smaller size.} This seems to
put current inflationary flow to considerable jeopardy.

In summary, the WMAP result of a varying spectral index, if
further stands, could be used as a discriminator for inflationary
models; most extant models would face a severe challenge. Using
the inflationary flow equations, we have studied in this paper the
{\bf possibility} of reconstructing inflation models with large
running spectral indices, which is favored by the WMAP analysis.

{\bf{Acknowledgments:}} Firstly we shall thank the anonymous
referees of this paper who inspired our thorough examination on
the flow scenario. We thank Profs. Zuhui Fan, Jun'ichi Yokoyama
and Xinmin Zhang for discussions and  Drs. Xuelei Chen, Mingzhe
Li, Bin Gong and Wanlei Guo for hospitable help. We acknowledge
the using of Numerical Recipes\cite{PressBK1}. This work was
supported in part by National Natural Science Foundation of China
under Grant No. 10273017  and by Ministry of Science and
Technology of China under Grant No. NKBRSF G19990754.

\bigskip
\bigskip
\bigskip

\bigskip
\noindent
{\footnotesize {\bf Table 1} -- Twelve sets of initial flow
parameters that lead to large running in the index as shown in
Fig.4.
\bigskip
\begin{center}
{\footnotesize
\begin{tabular}{|c|c|c|c|c|c|}
\hline
$\epsilon$   &$\delta$    &${}^2\lambda_{\rm H}$    &${}^3\lambda_{\rm H}$    &${}^4\lambda_{\rm H}$    &${}^5\lambda_{\rm H}$ \\
\hline
$0.10201$   &$    0.01683$ &    -0.03307 &   -0.02149  &   0.00575 &   $6.72964\times 10^{-4}$ \\
$0.18937$   &$   -0.341  $ &    -0.03109 &   -0.02458  &   0.00442 &   0.00183    \\
$0.15207$   &$   -0.26372 $&    -0.03191 &   -0.01817  &   0.00386 &   0.00128    \\
$0.07001$   &$   -0.4049  $&    -0.01344 &   -0.02067  &   0.01039 &  $-3.28768\times 10^{-5}$ \\
$0.14667$   &$   -0.47786 $&    -0.03289 &   -0.02483  &   0.00687&
0.00236 \\

$0.08029$   &$   -0.00671$ &    -0.04361 &   -0.01302  &   0.0068  &   $-4.26084\times 10^{-5}$\\

$0.049$   &$ -0.16761 $&   0.03082  & -0.02339  &  0.00379 &
$1.31775 \times 10^{-4}$ \\
$ 0.12419$ &$ -0.38872 $&   -0.01121 & -0.01841 &  $6.05911 \times
10^{-5}$ & 0.00274 \\
$0.1241$  &$  0.08762 $&  0.0374 &  -0.02455 & -0.00283 & 0.0017\\
$0.00532$  &$-0.35618 $& -0.00659 & 0.00402 & 0.0017 &
$1.77594\times 10^{-4}$ \\
$ 0.16525 $ &$ -0.39355 $&  -0.00507 & -0.01349 & $-6.855\times
10^{-4}$ & 0.00134\\
$0.16751$  &$ -0.40666 $&   -0.04769 &   -0.02294  &  0.00863 &  $7.67554\times 10^{-4}$\\

\hline
\end{tabular}
}
\end{center}
}


\begin{thebibliography}{nn}

\bibitem{kogut}
A. Kogut {\it et al.}, Astrophys. J. Suppl. 148, 161 (2003).

\bibitem{WMAP}\label{ref:map}
D. N. Spergel {\it et al.}, Astrophys. J. Suppl. 148, 175 (2003).

\bibitem{WMAP2}\label{ref:map2}
H. V. Peiris {\it et al.}, Astrophys. J. Suppl. 148, 213 (2003).

\bibitem{2df} W. J.~Percival {\it et al.}, Mon. Not. Roy. Astr. Soc.
{\bf 327}, 1297 (2001).

\bibitem{forest} R. A. C.~Croft {\it et al.}, Ap. J. {\bf 581}, 20 (2002);
N. Y.~Gnedin and A. J. S.~Hamilton, Mon. Not. Roy. Astr. Soc. {\bf
334}, 107 (2002).

\bibitem{Seljak} U. Seljak, P. McDonald and A. Makarov,
Mon.Not.Roy.Astron.Soc. 342, L79 (2003).

\bibitem{VSA}  S. Smith {\it et al.}, astro-ph/0401618;
R. Rebolo {\it et al.}, astro-ph/0402466; C. Dickinson {\it et
al.}, astro-ph/0402498;

\bibitem{feng}
B.~Feng, M.~Li, R.-J.~Zhang, and X.~Zhang, Phys. Rev. D {\bf 68},
103511 (2003).

\bibitem{running}J. E. Lidsey and R. Tavakol, Phys. Lett. B {\bf 575},
157 (2003); M. Kawasaki, M. Yamaguchi and J. Yokoyama, Phys.Rev. D
{\bf 68}, 023508 (2003); Q. G. Huang and M. Li, JHEP {\bf 0306},
014 (2003); D. J. Chung, G. Shiu and M. Trodden, Phys.Rev. D {\bf
68}, 063501 (2003); K.-I. Izawa, Phys. Lett. B {\bf 576}, 1
(2003); M. Bastero-Gil, K. Freese and L. Mersini-Houghton,
Phys.Rev.D {\bf 68} (2003) 123514; M. Yamaguchi and J. Yokoyama,
Phys.Rev.D {\bf 68} (2003) 123520; B. Wang, C. Lin and E. Abdalla,
Phys.Rev.D {\bf 69} (2004) 063507; G.Dvali and S. Kachru,
hep-ph/0310244; W. Lee, Y. Charng, D. Lee and L. Fang,
astro-ph/0401269; M. Yamaguchi and J. Yokoyama, hep-ph/0402282; D.
Lee, L. Fang, W. Lee and Y. Charng, astro-ph/0403055.


\bibitem{Lewis} S.~L.~Bridle, A.~M.~Lewis, J.~Weller, and G.~
Efstathiou, Mon.Not.Roy.Astron.Soc. 342, L72 (2003)


\bibitem{smalll}
e.g. C.~R.~Contaldi, M.~Peloso, L.~Kofman, and A.~Linde, JCAP {\bf
0307}, 002 (2003); E.~Gaztanaga et al, astro-ph/0304178;
J.~M.~Cline, P.~Crotty and J.~Lesgourgues, JCAP {\bf 0309} (2003)
002; B. Feng and X. Zhang, Phys.Lett.B {\bf 570}, 145 (2003); M.
Kawasaki and F. Takahashi, Phys.Lett.B {\bf 570}, 151 (2003); G.
Efstathiou, Mon.Not.Roy.Astron.Soc. {\bf 346}, L26 (2003); S.
Tsujikawa, R. Maartens and R. Brandenberger, Phys.Lett. B {\bf
574}, 141 (2003); T. Moroi and T. Takahashi, Phys.Rev.Lett. {\bf
92}, 091301 (2004); Q. Huang and M. Li, JCAP {\bf 0311} 001,
(2003); X. Bi, B. Feng and X. Zhang, hep-ph/0309195; J.-P. Luminet
et al, Nature {\bf 425} 593, (2003); Y. Piao, B. Feng and X.
Zhang, hep-th/0310206; Q. Huang and M. Li, astro-ph/0311378; L. R.
Abramo and L. Sodre, astro-ph/0312124; Y. Piao, S. Tsujikawa and
X. Zhang, hep-th/0312139; T. Multamaki and O. Elgaroy,
astro-ph/0312534.

\bibitem{Wang}P.~Mukherjee and Y.~Wang, Astrophys.J. 599,1 (2003).

\bibitem{fitWMAP}
V. Barger, H. Lee, and D. Marfatia, Phys.Lett.B {\bf 565}, 33
(2003);  S. M.Leach and A. R.Liddle, Phys.Rev.D {\bf 68} (2003)
123508.

\bibitem{Kinney03}
W. H.Kinney, E. W.Kolb, A. Melchiorri and A. Riotto,
hep-ph/0305130;


\bibitem{HT01}
M. B. Hoffman and M. S. Turner, Phys. Rev. D {\bf 64}, 023506
(2001).

\bibitem{kunz}
 S. H. Hansen and M. Kunz, Mon. Not. Roy. Astron. Soc. 336, 1007
 (2002).

\bibitem{Kinney02}
W. H. Kinney, Phys. Rev. D {\bf 66}, 083508 (2002).

\bibitem{Dodelson}
 S. Dodelson and L. Hui,  Phys.Rev.Lett.{\bf 91} 131301, (2003)

\bibitem{Ke03}
R. Easther and W. H. Kinney, Phys. Rev. D {\bf 67}, 043511 (2003).

\bibitem{liddle03flow}
A. R. Liddle, Phys. Rev. D {\bf 68}, 103504 (2003).

\bibitem{liddle94}
A. R. Liddle, P. Parsons, and J. D. Barrow, Phys. Rev. D {\bf 50},
7222 (1994).
\bibitem{stewart93}
E. D. Stewart and D. H. Lyth, Phys. Lett. B {\bf 302}, 171 (1993).

\bibitem{lidsey95}
J. E. Lidsey, A. R. Liddle, E. W. Kolb, E. J. Copeland and T.
Barreiro, Rev. Mod. Phys. {\bf 69}, 373 (1997).

\bibitem{fgw03}
B. Feng, X. Gong and X. Wang, astro-ph/0301111.

\bibitem{ckll03}
E. J. Copeland, E. W. Kolb, A. R. Liddle and J. E. Lidsey, Phys.
Rev. Lett. {\bf 71} 219 (1993).


\bibitem{linde91} A. D. Linde, Phys. Lett. {\bf 259B}, 38 (1991).

\bibitem{linde94} A. Linde, Phys. Rev. D {\bf 49} 748 (1994).

\bibitem{horder}
 e.g. D. H. Huang, W. B. Lin and X. M. Zhang, Phys. Rev. D 62,
087302 (2000); J. Martin and D. Schwarz, Phys. Rev. D 62, 103520
(2000); S. M. Leach, A. R. Liddle, J. Martin and D. J. Schwarz,
Phys. Rev. D 66, 023515 (2002); X. Wang {\it et al.},
astro-ph/0209242.

\bibitem{GStewart}
J. Gong and E. Stewart, Phys.Lett.B{\bf 510} 1, (2001).

\bibitem{STG01}
D. J. Schwarz, C. A. Terrero-Escalante and A. A. Garcia, Phys.
Lett. B 517, 243 (2001).

\bibitem{probflow}
W. H. Kinney, astro-ph/0307005.

\bibitem{PressBK1}
W. H. Press, S.~A. Teukolsky , W.~T. Vettering and B.~P. Flannery,
{\em Numerical Recipes in FORTRAN}, 2 ed. (Cambridge University
Press, Cambridge, 1992).

\end{thebibliography}
\end{document}